# Sharp selective scattering of red, green and blue light achieved via gain material's loss compensation

Yiyang Ye,[1*, a] Rongyue Liu,[1, a] Zhigang Song,[1] and T. P. Chen,[1*]

[1]*School of Electrical and Electronic Engineering, Nanyang Technological University, Singapore 639798, Singapore*
*a: The two authors contributed equally to this manuscript*
*\*E-mail: yeyi0005@e.ntu.edu.sg, echentp@ntu.edu.sg*

**Abstract**

For a transparent projection screen based on metallic nanoparticle's localized surface plasmon resonance (LSPR), in the ideal case the metallic nanoparticles dispersed in a transparent matrix only selectively scatter red, green and blue light and transmit the visible light of other colours. However, metal's optical loss and size effect at nanoscale degenerate the desired performance by broadening the resonance peak width and increasing the absorption ratio. In this work, it is shown that the problem can be solved by introducing gain material. Numerical simulations are performed on nanostructures based on silver (Ag), gold (Au) or aluminium (Al) with or without gain material, to examine the effect of gain material and to search for suitable structures for sharp selective scattering of red, green and blue light. And it is found that introducing gain material greatly improves performance of the structures based on Ag or Au except the structures based on Al. The most suitable structures for sharp selective scattering of red, green and blue light are respectively found to be the core-shell structures of silica@Au, silica@Ag and Ag@silica, all with gain material.

**Introduction**

Recently, the idea of wavelength-selective scattering of light to achieve transparent projection screen has been proposed based on metallic nanoparticle's localized surface plasmon resonance (LSPR),[1-4] where in the ideal case metallic nanoparticles dispersed in a transparent matrix only selectively scatter red, green and blue light and transmit visible light of other colours. Nevertheless, due to metal's optical loss, light absorption is inevitable in the LSPR process of the metallic nanoparticles, which is undesirable for scattering-based applications. In addition, at nanoscale, surface scattering of conduction electrons becomes significant,[5] which degenerates the quality of selective scattering based on LSPR by further increasing absorption ratio as well as broadening the resonance peak width.[6-10] Yet, these problems may be solved by incorporating gain materials such as rare earth ions and laser dyes, which compensate optical loss, narrow LSPR peak width and are even theoretically demonstrated to enable lasing state. [11-20]

In this work, through numerical simulation, we aim to achieve sharp scattering peaks at red, green and blue light, with absorption being maintained low over the visible light range (400 nm ~ 800 nm) based on metallic nanoparticle's LSPR effect and gain material's loss compensation. Optimizations based on metallic nanosphere and core-shell structures of metal@silica and silica@metal are conducted for gold (Au), silver (Ag) and aluminium (Al), with and without gain material, with the size effect of metallic nanoparticle on metal's dielectric function taken into account. The results from the optimizations show that for the Ag- and Au-based structures, introducing gain material significantly improves the scattering behaviour (i.e., there are sharper scattering peaks at red, green and blue light, respectively, and low absorption over visible light range), however the improvement is not significant for the Al-based structures. And it is found that the most suitable structures for sharp selective scattering at the red, green and blue light wavelength regions are silica@Au, Ag@silica and silica@Ag, respectively, all with a gain material.

**Simulation method**

In this section, reasons of choosing the metals and structures for subsequent optimizations are given, and the method of optimization is described.

Since in this work, selective scattering of light is based on metallic nanoparticle's LSPR, and LSPR is a result of resonant oscillation of free electrons in response to incident electromagnetic wave,[21-22] free electron behaviours of metals to be employed in subsequent simulations should be clearly separated from the broad spectrums caused by inter-band transitions of bound electrons.[23] According to this criteria, alkali metals, Al, Au and Ag are the candidates. But alkali metals are excluded because they are very reactive. So, only Al, Au and Ag will be used in the following simulations.

The surrounding medium's refractive index $n_s$ is assumed to be a constant of 1.5, which is about the same as the refractive indices of glass and most polymers, in which the nanoparticles to be designed are supposed to be dispersed. In quasi-static limit, namely, when the size of a metallic nanosphere is much smaller than the wavelength of incident light, the LSPR peak appears at a particular wavelength at which the real part of the metal's dielectric function is equal to negative 2 times of surrounding medium's dielectric function,[24] i.e. $Re[\varepsilon_m(\lambda_R)] = -2 \cdot \varepsilon_s$, where $\lambda_R$ is the resonance peak wavelength and $\varepsilon_m(\lambda_R)$ is metal's dielectric function at this wavelength, and $\varepsilon_s (= n_s{}^2)$ is surrounding medium's dielectric function. For Ag, Au and Al, in quasi-static limit, their resonance peaks appear at 416 nm, 540 nm and 192 nm respectively, as observed from Figure 1 below.

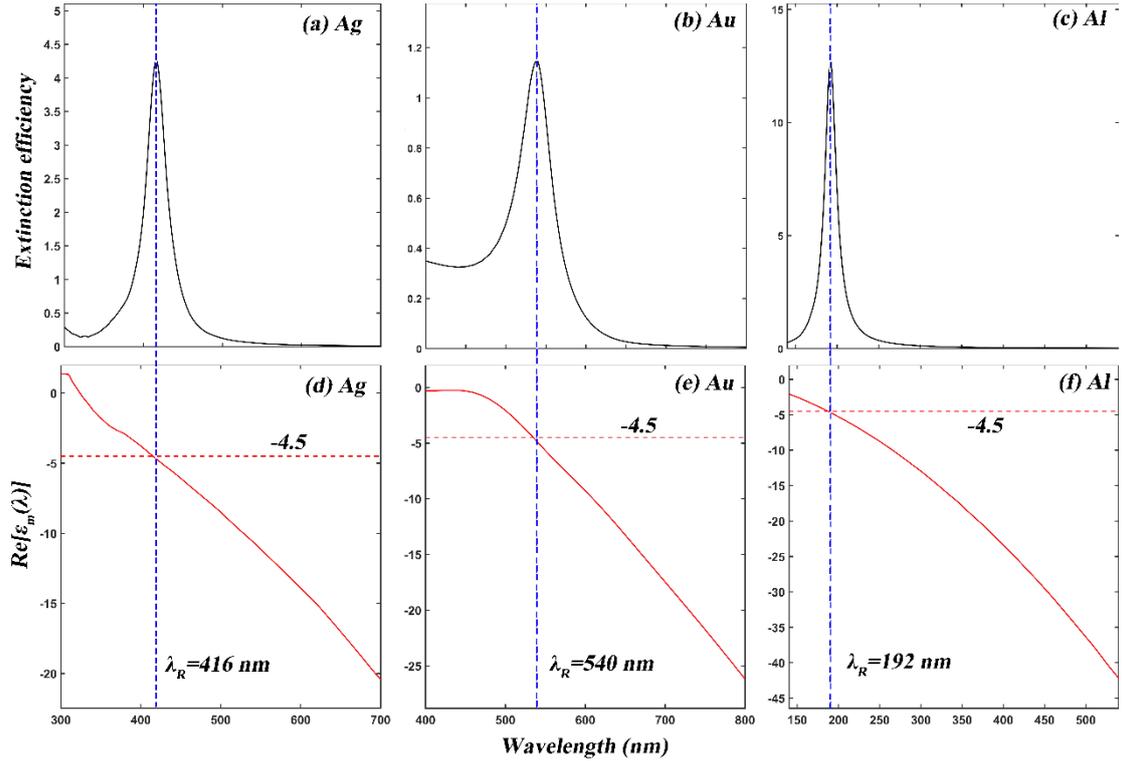

*Figure 1. Calculated extinction efficiency spectrum of a single nanosphere with diameter of 10 nm by Mie's theory,[23] for (a) Ag, (b) Au and (c) Al. The real part of dielectric function is plotted for (d) Ag, (e) Au and (f) Al. The surrounding medium's refractive index is 1.5, so surrounding medium's dielectric function is $\varepsilon_s = n_s^2 = 2.25$, and the resonance peaks appear at the wavelengths at which $Re[\varepsilon_m(\lambda)] = -2 \cdot \varepsilon_s = -4.5$. The dielectric functions of the metals are from literature.[25]*

Increasing metallic nanosphere's size can red-shift its resonance peak, but further decreasing the size of a nanosphere with a size much smaller than the wavelength of incident light does not significantly blue-shift its resonance peak.[24] Thus, here, the resonance peak wavelengths for Ag, Au and Al spheres in the quasi-static limit are perceived as the lower limits of their corresponding tuning ranges of resonance peak wavelength. The wavelengths of red, green and blue light are around 630 nm, 520 nm and 450 nm respectively. So, here, Au is considered only for red light scattering, as the lower limit (540 nm) of its resonance-peak-wavelength tuning range is larger than the wavelength of green light, while Ag and Al are considered for scattering of all three colours of light. The next step is to red-shift the metallic nanoparticles' resonance peak wavelengths from their respective ones in the quasi-static limit, to red light for Au, and to all three colours of light for Ag and Al. A small red-shift of resonance peak of less than about 50 nm can be achieved by increasing the metallic nanospheres' size. But relying on increasing size to red-shift resonance peak by more than 100 nm has the following problem: the dipole resonance peak's strength will decrease, and the quadruple mode peak will appear and gradually become dominant, resulting in an overall broad extinction spectrum.[24] A feasible solution to achieve flexible red-shift of resonance peak is to use the structure of metallic nanoshell, which easily tunes the resonance peak wavelength via the ratio of shell thickness to core diameter.[1, 10, 26] Therefore, the structures of metallic nanosphere and nanoshell are employed in subsequent optimizations when gain material is not included, with the assumption that the dielectric core of nanoshell is silica. In the case of metallic nanoshell, the gain material is introduced by doping the silica core with gain material; while in the metallic nanosphere's case, the gain material is introduced by surrounding the metallic nanosphere by a shell of silica doped with gain material.

With the above considerations, the structures to be considered during optimization are summarized in Table 1 below.

*Table 1* Structures on which optimizations are to be performed to search for solutions of sharp selective scattering of red, green and blue light.

|  | Without gain material | With gain material |
|---|---|---|
| Red | Au sphere, silica@Au, silica@Ag, silica@Al | Au@silica, silica@Au, silica@Ag, silica@Al |
| Green | Ag sphere, silica@Ag, silica@Al | Ag@silica, silica@Ag, silica@Al |
| Blue | Ag sphere, silica@Ag, silica@Al | Ag@silica, silica@Ag, silica@Al |

For the purpose of optimization, a figure of merit ($FOM$) is defined:

$$FOM = \frac{\sigma_{sca}(\lambda_R)}{2\overline{\sigma_{sca}} + \max\{|\sigma_{abs}|\}} \tag{1}$$

where $\sigma_{sca}$ and $\sigma_{abs}$ are the scattering and absorption cross sections, respectively which are calculated by the Mie's theory,[23] or the extension of the theory for core-shell structure.[27] The overbar and the symbol $max\{...\}$ denote the mean and the maximum values respectively in the visible spectrum (from 400 nm to 800 nm), and $\lambda_R$ is the resonance wavelength as mentioned earlier. This $FOM$ is defined in the way reported in reference,[1] except the introduction of an absolute value operator on $\sigma_{abs}$. This $FOM$ favours the following properties: high scattering cross section at resonant wavelength and low scattering cross section elsewhere to form a sharp scattering peak and an overall small value of absorption cross section. Introducing gain material can make the absorption cross section negative, and eventually result in a lasing state[16] if the absolute value operator is not put on $\sigma_{abs}$ in equation (1) during optimization of $FOM$. The purpose of introducing gain material is to just compensate light absorption caused by metallic nanoparticle's optical loss, and a low magnitude of overall absorption cross section is preferred over its mathematical value.

Since at nanoscale, the size of a metallic nanoparticle is comparable to free electrons' mean free path in the bulk metal, effect of surface scattering of conduction electrons on the metal's dielectric function cannot be ignored.[5] This issue is taken into account in equation (2):[28]

$$\varepsilon_m(\omega) = \varepsilon_{exp}(\omega) + \frac{\omega_p^2}{\omega(\omega + i\gamma_b)} - \frac{\omega_p^2}{\omega[\omega + i(\gamma_b + \gamma_s)]} \tag{2}$$

where $\varepsilon_m$ is metal's dielectric function corrected for size effect, $\omega$ is frequency of incident light, $\varepsilon_{exp}$ is the dielectric function of the bulk metal (Ag, Au or Al) obtained from literature,[25] $\omega_p$ and $\gamma_b$ are the bulk metal's plasma frequency and damping rate, respectively, whose values for Ag, Au and Al are given in Table 2[28], $i$ is the imaginary number, $\gamma_s$ is the extra damping of free electrons caused by surface scattering and is given by equation (3):[28]

$$\gamma_s = \frac{v_F}{L_B} \tag{3}$$

where $v_F$ is the Fermi velocity of electrons in metal, and its values for Ag, Au and Al are given in Table 2,[28] and $L_B$ is the effective mean free path of free electrons in metallic nanoparticle. For the structure of metallic nanoshell, i.e., silica@metal, $L_B$ is given by equation (4), derived by assuming specular reflections of electrons from surfaces of metallic shell:[29]

$$L_B = \frac{4(r_o^3 - r_i^3)}{3(r_o^2 + r_i^2)} \tag{4}$$

where $r_o$ and $r_i$ are the outer and inner radii of the core-shell structure, respectively. For metallic nanosphere or the silica-coated metallic nanosphere (i.e., metal@silica), $L_B$ is given by equation (5), which is derived by setting $r_i$ to 0 in equation (4):

$$L_B = \frac{4}{3}r \tag{5}$$

where $r$ is the radius of metallic nanosphere or the inner radius of silica-coated metallic nanosphere.

*Table 2 Plasma frequencies, damping rates and Fermi velocities of bulk metals used in correcting the metals' dielectric functions for the size effect.*[28]

|    | $\omega_p$(eV) | $\gamma_b$(eV) | $v_F(10^6 m/s)$ |
|----|----------------|----------------|-----------------|
| Ag | 9.6            | 0.0228         | 1.39            |
| Au | 8.55           | 0.0184         | 1.40            |
| Al | 15.3           | 0.5984         | 2.03            |

For silica without gain material, it is assumed to have a constant refractive index $N_{Si} = 1.45$. For silica doped with gain material, its refractive index is given by[16]

$$N_{Si} = 1.45 + ik \tag{6}$$

where $i$ is the imaginary number, and $k$ is a negative extinction coefficient constant over all wavelengths, since reported gain materials are able to cover the whole visible wavelength range.[11, 13] Extinction coefficient $k$ is positive for lossy media, 0 for non-absorbing media, and negative for gain media.

The optimization is carried out with BOBYQA,[30] a local optimization algorithm, which starts from an initial guess of tuning parameters and searches for an optimized set of tuning parameters that give maximum value of $FOM$. Here, the set of tuning parameters are dimensions of the nanostructures shown in Table 1 (i.e., core diameter and shell thickness for metal@silica and silica@metal, or diameter for metal sphere), the value of $k$ in equation (6), and resonance peak wavelength $\lambda_R$. The tuning ranges of resonance peak wavelength $\lambda_R$ for red, green and blue light are set to be 620 nm to 700 nm, 510 nm to 540 nm, and 440 nm to 460 nm, respectively. The detailed optimization process is as following: for each tuning range of $\lambda_R$, $\lambda_R$ is scanned at a step of 1 nm; at each step of $\lambda_R$, BOBYQA is carried out with an initial guess of nanostructure's dimensions and $k$ value; and the set that generates the largest value of $FOM$ defined by equation (1) gives the optimized $\lambda_R$, nanostructure's dimensions and $k$ value. For structures without gain material, $k$ equals to 0 and is not an adjustable parameter during optimization. The optimized results of nanostructure's dimensions, $k$ values and values of optimized $FOM$ are shown in the section of "Results and discussion".

**Results and discussion**
In this section, apart from the optimized tuning parameters, and values of optimized $FOM$, the calculated optical spectrums are also presented to give an intuitive view of sharp selective scattering.

The spectrums are presented in terms of extinction efficiency ($Q_{ext}$), scattering efficiency ($Q_{sca}$) and absorption efficiency ($Q_{abs}$), which are obtained from dividing their corresponding cross sections: $\sigma_{ext}$, $\sigma_{sca}$ and $\sigma_{abs}$, calculated by Mie's theory,[23,27] to the geometrical cross section $\pi R^2$, where $R$ is the overall radius of a core-shell structure (metal@silica or silica@metal) or radius of a nanosphere, i.e., $Q_{ext} = \sigma_{ext}/(\pi R^2)$, $Q_{sca} = \sigma_{sca}/(\pi R^2)$, and $Q_{abs} = \sigma_{abs}/(\pi R^2)$. Note that $Q_{ext}$ is equal to the sum of $Q_{sca}$ and $Q_{abs}$.

For the calculated efficiencies shown in Figures 2, 3 and 4, Mie's theory takes the following parameters as its inputs: the surrounding medium's dielectric function $\varepsilon_s = n_s^2 = 2.25$, optimized nanostructure's dimensions (i.e., core diameter and shell thickness for a core-shell structure, or diameter for a nanosphere), metal's size-dependent dielectric function given by equation (2), silica's dielectric function $\varepsilon_{Si}$ for the core-shell structure (for a structure without gain material, $\varepsilon_{Si} = N_{Si}^2 = 2.1025$; for a structure with gain material $\varepsilon_{Si} = N_{Si}^2 = (1.45 + ik)^2$, where $N_{Si} = 1.45 + ik$ is from equation (6) and the optimized $k$ is constant over the whole spectrum).

Since one gain material only provides gain in a small wavelength range, a negative $k$ over a short wavelength range is more physically meaningful. For this purpose, for those structures with gain material, we have also investigated the effect on value of $FOM$ when replacing the wavelength-independent $k$ with a gaussian-form $k$. Here,

the gaussian-form $k$ has the following expression:

$$k(\lambda) = k_{optimal} e^{-\frac{(\lambda-\lambda_R)^2}{2a^2}} \tag{7}$$

where $k_{optimal}$ and $\lambda_R$ are the optimized $k$ value and resonance wavelength respectively for a structure with gain material, $\lambda$ is incident light's wavelength, $a$ is associated with gaussian curve's full width at half maximum (FWHM), and FWHM is assumed to be 30 nm in this work. The gaussian-form $k$ is equal to the optimized $k$ in Tables 3, 4 and 5 at the resonance wavelength, and decreases in magnitude for wavelengths away from the resonance wavelength. For structures with gain material, their values of $FOM$ calculated with gaussian-form $k$, with other optimized parameters unchanged, are also shown in Tables 3, 4 and 5, and their efficiency spectrums calculated with gaussian-form $k$ are shown in Figures 5, 6 and 7. Nevertheless, $k$ is assumed to be wavelength-independent during optimization and in Figures 2, 3 and 4.

Figure 2 shows the efficiency spectrums of the structures optimized for selective scattering of red light, and their corresponding values of $FOM$ as well as optimized parameters are shown in Table 3.

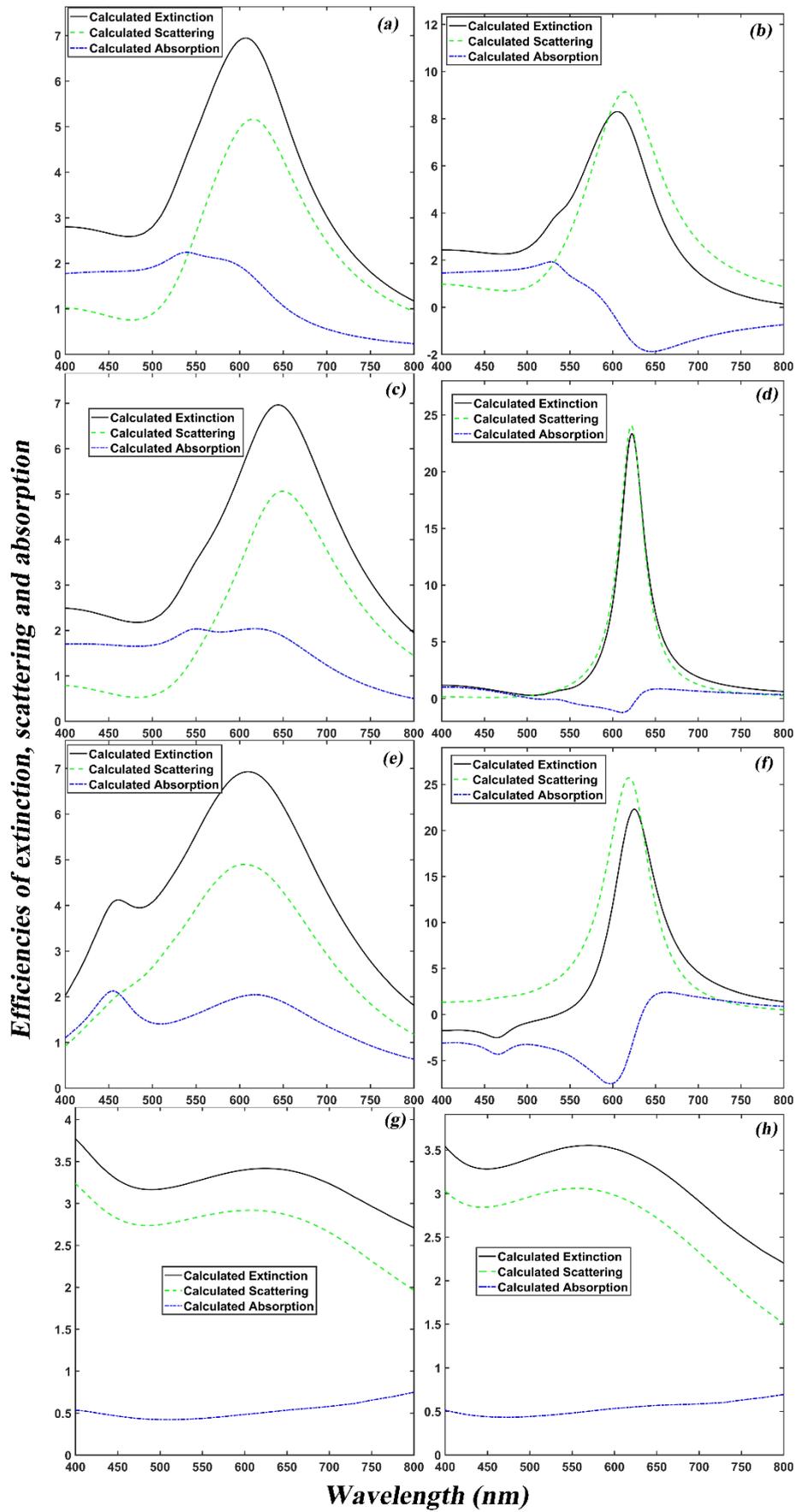

*Figure 2. Efficiency spectrums of optimized structures for selective scattering of red*

*light. The left column, i.e., (a), (c), (e) and (g) are for the optimized structures without gain material, and spectrums of their corresponding structures optimized with gain material are shown in the right column, i.e., (b), (d), (f) and (h). (a) Au sphere without gain. (b) Au@silica with gain. (c) Silica@Au without gain. (d) Silica@Au with gain. (e) Silica@Ag without gain. (f) Silica@Ag with gain. (g) Silica@Al without gain. (h) Silica@Al with gain. For each structure, the optimized parameters and optimized values of $FOM$ are tabulated in Table 3 in the same alphabetic order as their respective structure appears in this figure.*

*Table 3 Optimized tuning parameters and values of optimized $FOM$ for the structures mentioned in Figure 2, with the same alphabetic order*

| Alphabetic order | Core Diameter (nm) | Shell thickness (nm) | $k$ | Resonance wavelength (nm) | $FOM$ | $FOM$ with Gaussian $k$ |
|---|---|---|---|---|---|---|
| (a) | 101 | | | 620 | 0.7476 | |
| (b) | 98 | 5 | -0.4663 | 620 | 1.1094 | 1.4925 |
| (c) | 61 | 23 | | 650 | 0.7640 | |
| (d) | 48 | 10 | -1.2888 | 622 | 3.17 | 2.7835 |
| (e) | 65 | 18 | | 620 | 0.6063 | |
| (f) | 56 | 8 | -1.1234 | 620 | 1.3391 | 3.1377 |
| (g) | 97 | 25 | | 672 | 0.4509 | |
| (h) | 83 | 25 | -0.0677 | 622 | 0.4853 | 0.4856 |

Figure 3 shows the efficiency spectrums of the structures optimized for selective scattering of green light, and their corresponding value of $FOM$ as well as optimized parameters are shown in Table 4.

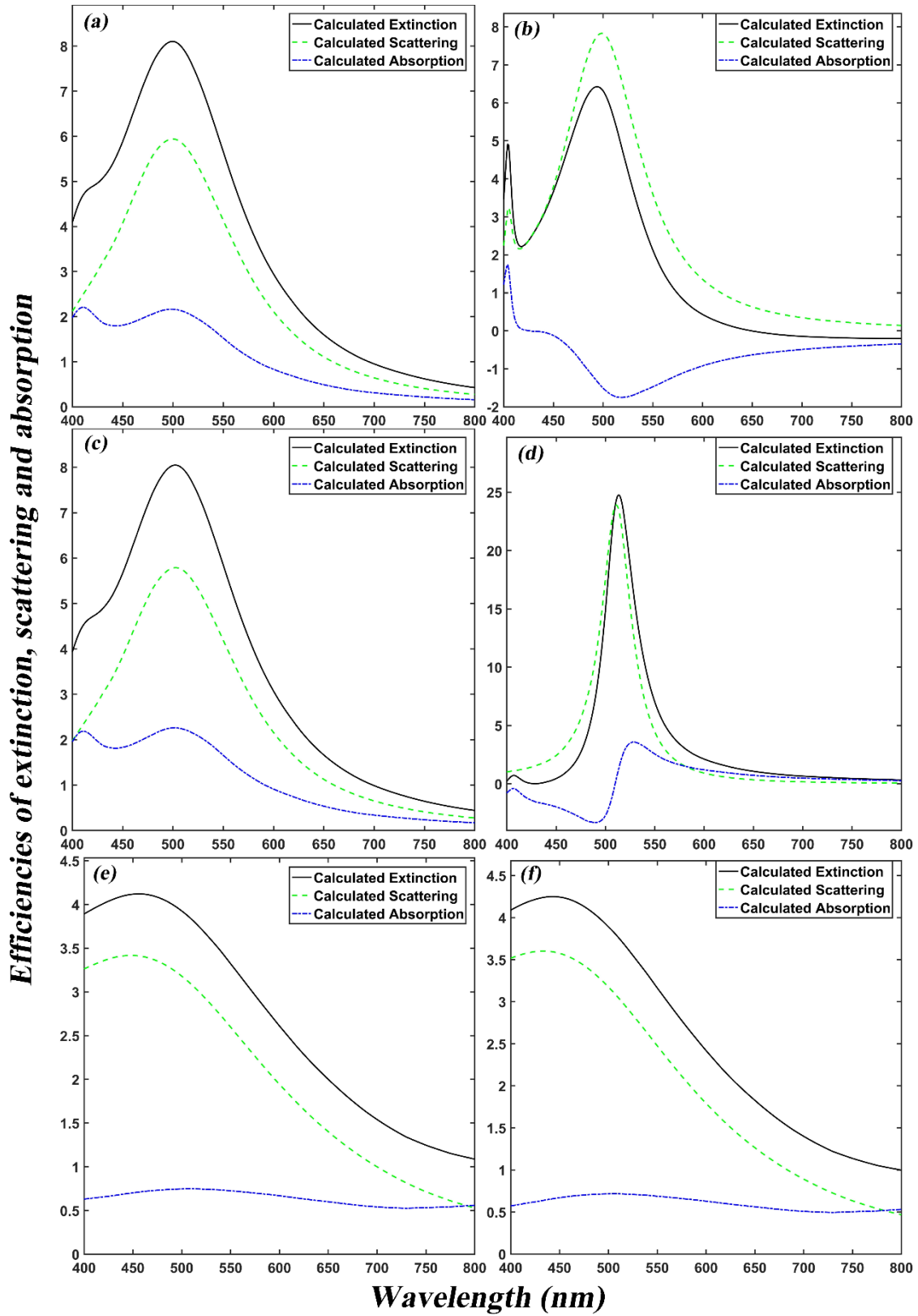

*Figure 3. Efficiency spectrums of optimized structures for selective scattering of green light. The left column, i.e., (a), (c) and (e) are for the optimized structures without gain material, and spectrums of their corresponding structures optimized with gain material are shown in the right column, i.e., (b), (d) and (f). (a) Ag sphere without gain. (b)*

*Ag@silica with gain. (c) Silica@Ag without gain. (d) Silica@Ag with gain. (e) Silica@Al without gain. (f) Silica@Al with gain. For each structure, the optimized parameters and optimized values of FOM are tabulated in Table 4 in the same alphabetic order as their respective structure appears in this figure.*

*Table 4 Optimized tuning parameters and values of optimized $FOM$ for the structures mentioned in Figure 3, with the same alphabetic order*

| Alphabetic order | Core Diameter (nm) | Shell thickness (nm) | $k$ | Resonance wavelength (nm) | $FOM$ | $FOM$ with Gaussian $k$ |
|---|---|---|---|---|---|---|
| (a) | 78 | | | 510 | 0.8163 | |
| (b) | 78 | 15 | -0.1743 | 510 | 1.1289 | 1.58 |
| (c) | 20 | 29 | | 510 | 0.8087 | |
| (d) | 34 | 9 | -1.2435 | 511 | 2.4128 | 2.6309 |
| (e) | 65 | 18 | | 510 | 0.6329 | |
| (f) | 62 | 18 | -0.8485 | 510 | 0.6522 | 0.6597 |

Figure 4 shows the efficiency spectrums of the structures optimized for selective scattering of green light, and their corresponding value of $FOM$ as well as optimized parameters are shown in Table 5.

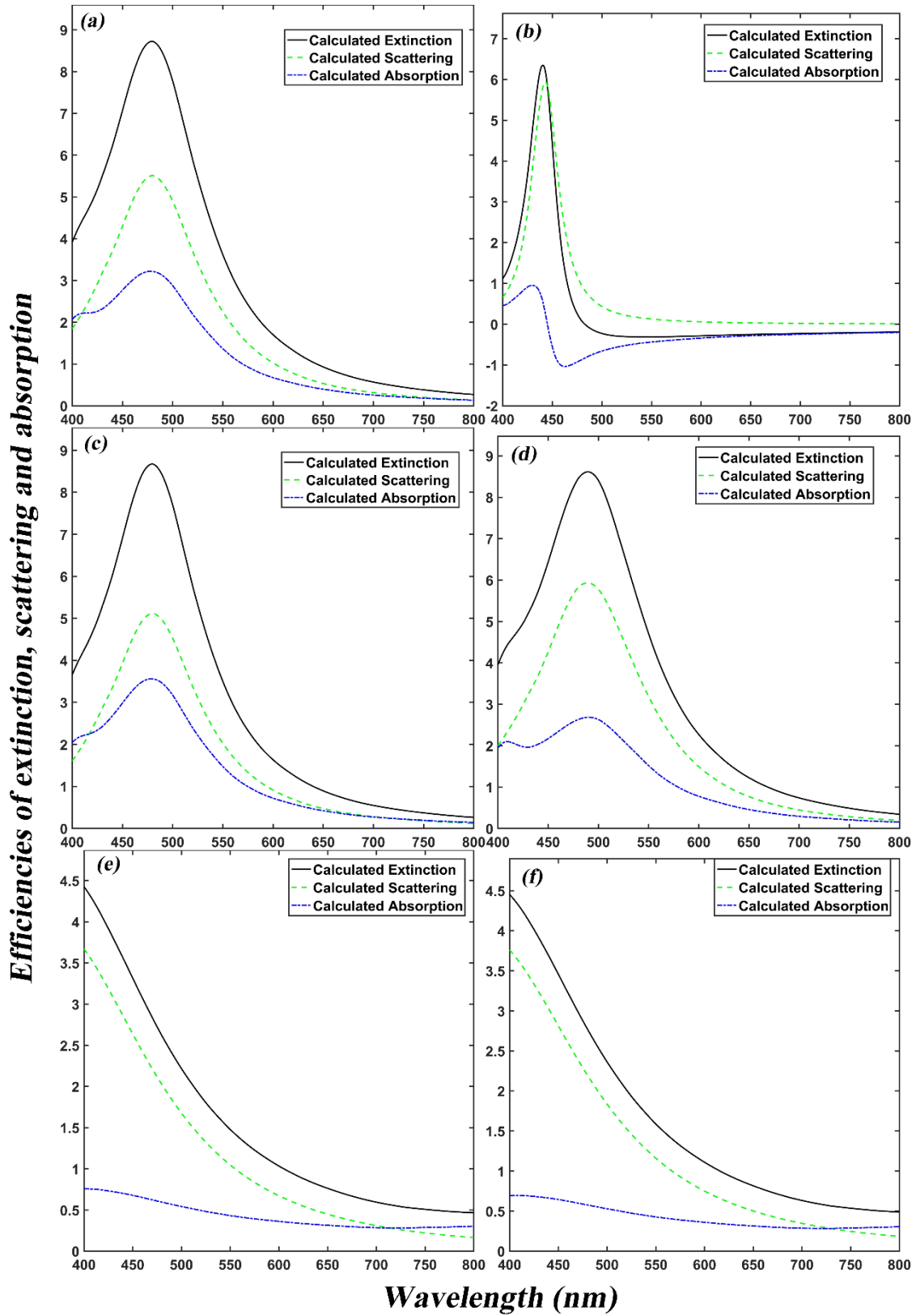

*Figure 4. Efficiency spectrums of optimized structures for selective scattering of blue light. The left column, i.e., (a), (c) and (e) are the optimized structures without gain material, and spectrums of their corresponding structures optimized with gain material are shown in the right panels, i.e., (b), (d) and (f). (a) Ag sphere without gain. (b)*

Ag@silica with gain. (c) Silica@Ag without gain. (d) Silica@Ag with gain. (e) Silica@Al without gain. (f) Silica@Al with gain. For each structure, the optimized parameters and optimized values of $FOM$ are tabulated in Table 5 in the same alphabetic order as their respective structure appears in this figure.

*Table 5* Optimized tuning parameters and values of optimized $FOM$ for the structures mentioned in Figure 4, with the same alphabetic order

| Alphabetic order | Core Diameter (nm) | Shell thickness (nm) | $k$ | Resonance wavelength (nm) | $FOM$ | $FOM$ with Gaussian $k$ |
|---|---|---|---|---|---|---|
| (a) | 67 | | | 460 | 0.7123 | |
| (b) | 47 | 19 | -0.1726 | 443 | 2.5227 | 2.4159 |
| (c) | 17 | 24 | | 460 | 0.6569 | |
| (d) | 18 | 27 | -1.44 | 460 | 0.6923 | 0.6960 |
| (e) | 31 | 23 | | 440 | 0.9602 | |
| (f) | 27 | 26 | -0.8688 | 440 | 0.9794 | 0.9780 |

For the structures with gain material in Figures 2, 3 and 4, their spectrums are calculated assuming the optimized values of $k$ to be wavelength-independent, and Figures 5, 6 and 7 below show the calculated spectrums with the gaussian-form k.

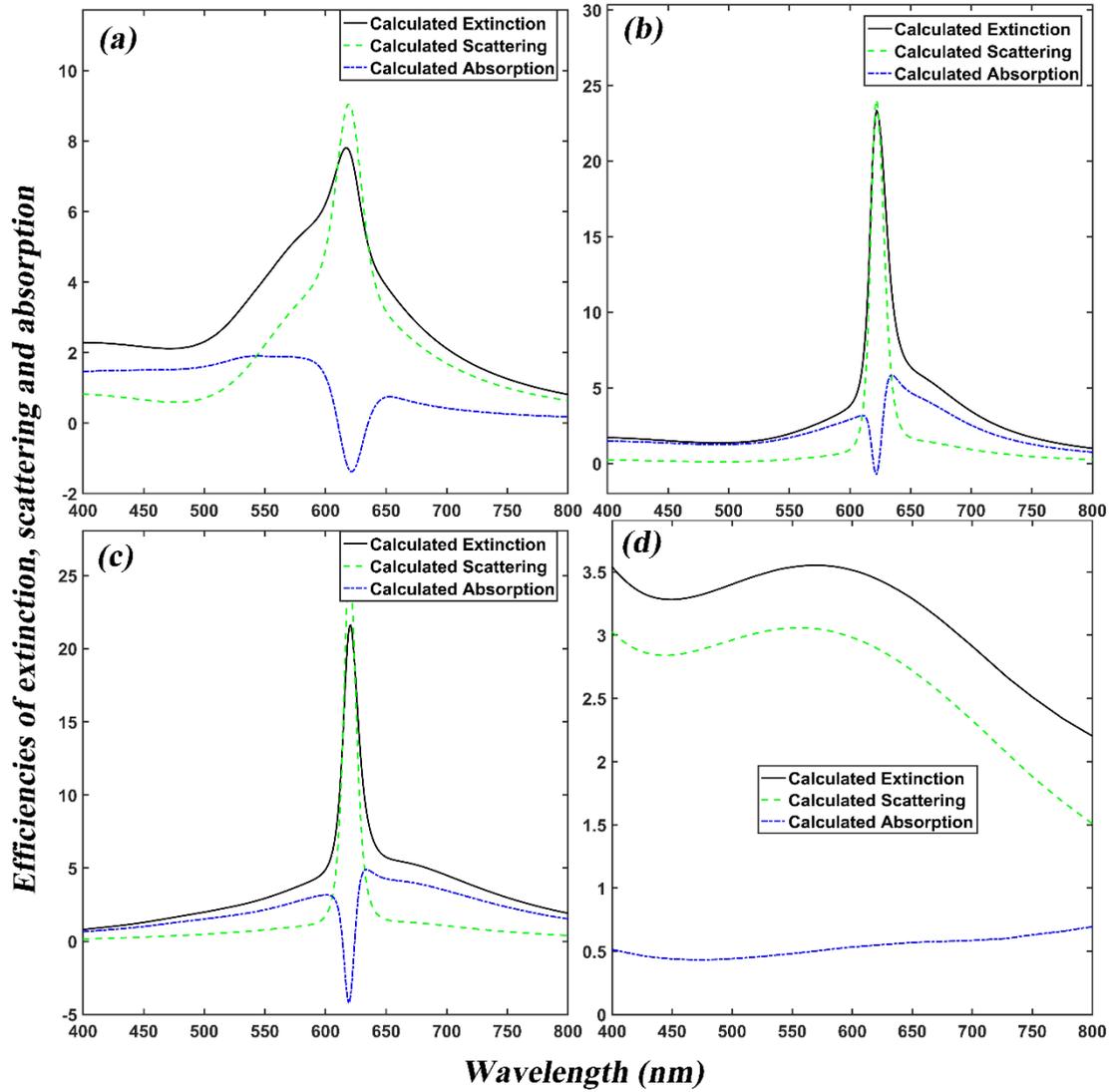

*Figure 5.* *Efficiency spectrums recalculated for structures with gain material in Figure 2, all parameters used in calculation here are same as those used in Figure 2 except that the wavelength-independent k in Figure 2 is replaced with gaussian-form k here. (a) Au@silica with gain. (b) Silica@Au with gain. (c) Silica@Ag with gain. (d) Silica@Al with gain. Structures of (a), (b), (c) and (d) in this figure correspond to (b), (d), (f) and (h) in Figure 2, respectively.*

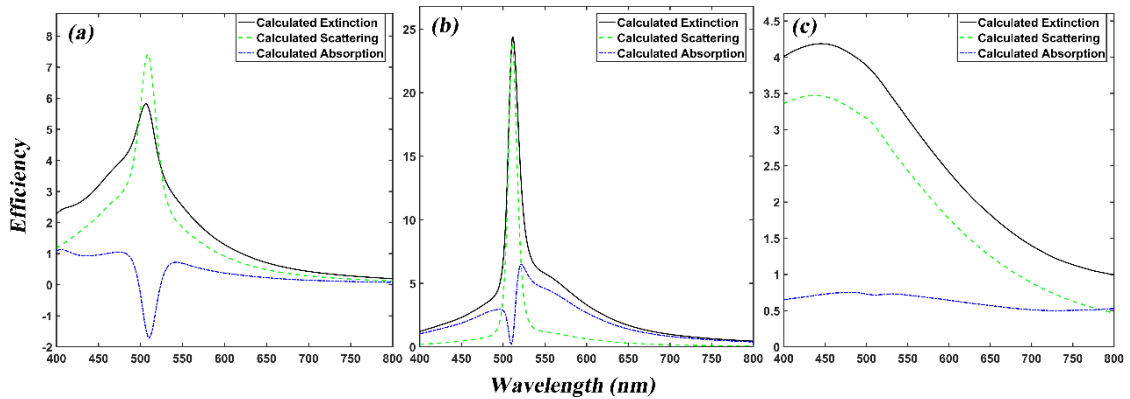

*Figure 6.* Efficiency spectrums recalculated for structures with gain material in Figure 3, all parameters used in calculation here are same as those used in Figure 3 except that the wavelength-independent $k$ in Figure 3 is replaced with gaussian-form $k$ here. (a) Ag@silica with gain. (b) Silica@Ag with gain. (c) Silica@Al with gain. Structures of (a), (b) and (c) in this figure correspond to (b), (d) and (f) in Figure 3, respectively.

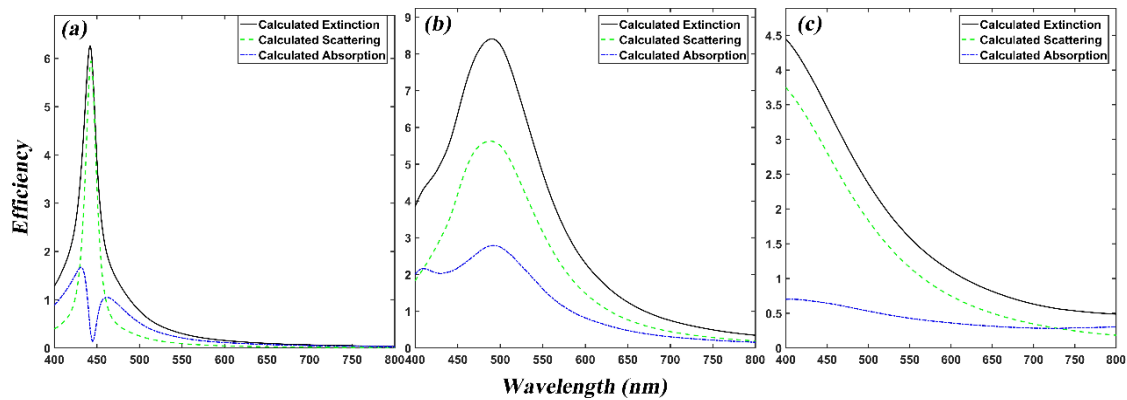

*Figure 7.* Efficiency spectrums recalculated for structures with gain material in Figure 4, all parameters used in calculation here are same as those used in Figure 4 except that the wavelength-independent $k$ in Figure 4 is replaced with gaussian-form $k$ here. (a) Ag@silica with gain. (b) Silica@Ag with gain. (c) Silica@Al with gain. Structures of (a), (b) and (c) in this figure correspond to (b), (d) and (f) in Figure 4, respectively.

Negative absorption is observed for some structures, which is interpreted as radiation of light. The occurrence of negative absorption is reasonable with the introduction of gain material, as gain material is used in laser to achieve lasing state which is a state of light radiation.

By observing Figures 2, 3 and 4, it is obvious that Al-based structures are not suitable for selective scattering, since their resonance peaks are too broad, even though absorptions are low. The efficiency spectrums for Al-based structures do not show apparent improvements even after gain material is introduced. And the broad resonance peaks for Al-based structures can be explained by Al's low quality factor, which determines strength of resonance, defined by $-Re[\varepsilon_m]/Im[\varepsilon_m]$, over the whole visible wavelength range.[28]

Apart from structures of Al, by comparing the calculated efficiency spectrums of the structures without gain material with those of structures with gain material in Figures 2, 3 and 4, it is observed that introducing gain material significantly improves the scattering quality in the sense that the ratios of absorption are decreased, and the resonance scattering peaks become sharper. And the improvements after introducing gain material are confirmed by the increase of values of $FOM$ as shown in Tables 3, 4 and 5.

The largest values of $FOM$ in Tables 3, 4 and 5 (3.17, 2.4128 and 2.5227 respectively) suggest that the most suitable structures for sharp selective scattering of red, green and blue light are the structures of silica@Au with gain material (see Table 3), silica@Ag with gain material (see Table 4), and Ag@silica with gain material (see Table 5), respectively. Their angular distributions of scattering are shown in Figure 8. The results shown in Figure 8 suggest that the scattered light can be viewed from a wide range of angle. The $FOM$ values of the most suitable structures (i.e., 3.17 for red light scattering, 2.4128 for green light scattering, and 2.5227 for blue light scattering) are much larger than that reported in previous works,[1, 3-4] where a $FOM$ value of 1.5 is considered high, showing a great improvement in selective scattering of red, green and blue light.

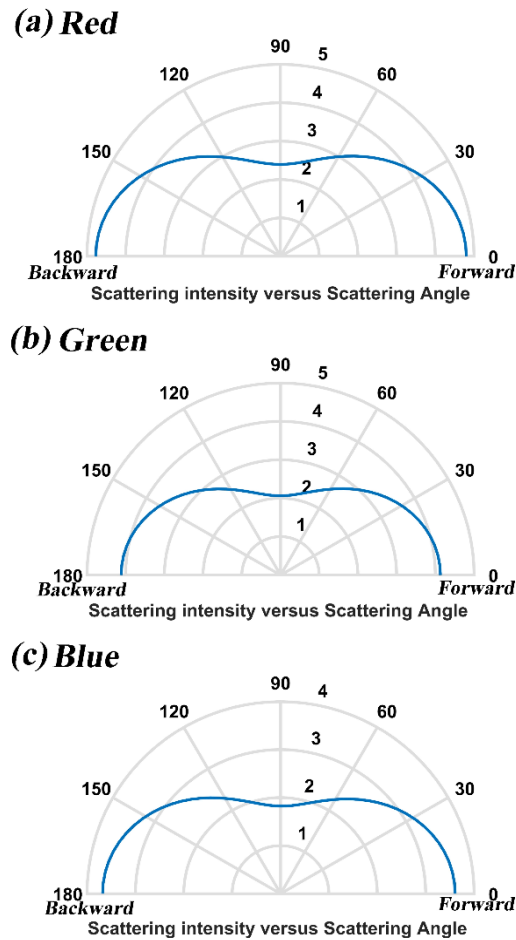

*Figure 8. Intensity of the scattered light at the resonant wavelengths versus the scattering angle (in the unit of degree) in polar form for the most suitable structures respectively for selective scattering of red, green and blue light: (a) Silica@Au with gain material (Table 3), (b) Silica@Ag with gain material (Table 4), and (c) Ag@silica with gain material (Table 5). The incident light has a unit intensity and equal intensity for the p- and s- components with respect to the scattering plane. Calculations are carried out with the Mie's theory.*

For the structures with gain material, after replacing wavelength-independent $k$ with gaussian-form $k$, change in $FOM$ is not significant except for the structure of Silica@Ag with gain material (item f in Table 3). For this structure with gaussian-form $k$, $FOM$ increases by a factor of 2.34 resulting in a value of 3.1377, and is almost the same as the largest $FOM$ of 3.17 in Table 3. This suggests that this structure with gaussian-form $k$ is an alternative structure for selective scattering of red light.

Among the three most suitable structures, the one for selective scattering of red light, i.e., the structure of silica@Au with gain material (item d in Table 3), has the largest magnitude of optimized $k$ which is 1.2888, and this corresponds to a gain value of $2.6 \times 10^5 cm^{-1}$, according to the relation $gain = -\frac{4\pi k}{\lambda}$,[11] where $\lambda = 622 nm$ is this structure's optimized resonance wavelength. The relation between gain and the emission cross section $\sigma_e$ of a single gain material (dye or rare earth ion) is: $gain = \rho \cdot \sigma_e$,[14] where $\rho$ is gain material's volume concentration. The usually-known gain material's emission cross section is of the order of $10^{-16} cm^2$,[11-12, 14] which corresponds to a maximum gain of the order of $10^3 cm^{-1}$. However, it has been pointed out that the theoretical limit of emission cross section set by quantum mechanics is $3\lambda^2/(2\pi)$,[31] which is of the order of $10^{-10} cm^2$ for visible wavelength range, and is 6 orders larger than $10^{-16} cm^2$. So, a gain value of the order of $10^5 cm^{-1}$ may be achieved theoretically.

**Conclusion**
In this work, for the purpose of realizing a transparent projection screen, we have searched for structures that may achieve sharp selective scattering of red, green and blue light based on metallic nanoparticle's LSPR effect and gain material's optical compensation. By comparing behaviors of structures optimized with gain material to those without gain material, it is found that for the structures based on Au and Ag, introducing gain material greatly improves the performance in terms of sharper scattering peak and overall lower absorption, while for the Al-based structures little improvement is observed. The most suitable structures for sharp selective scattering of red, green and blue light are respectively found to be silica@Au, silica@Ag and Ag@silica, all with gain material, with their corresponding optimized parameters shown in Tables 3, 4 and 5. And among gain values of the three most suitable structures, the largest one is still within physical limit.

**Acknowledgements**
This work was financially supported by the National Research Foundation of Singapore (Program Grant No. NRF-CRP13-2014-02).